\documentclass[showkeys,showpacs]{revtex4}
\usepackage{amsmath}
\usepackage{amssymb}
\usepackage{graphicx,color}
\usepackage[cp1251]{inputenc}
\usepackage[russian]{babel}

\begin{document}

\title{Blind physics: a catalogue of unverified hypotheses}

\author{Vladimir Dzhunushaliev}
\email{vdzhunus@krsu.edu.kg}
\affiliation{
Institute for Basic Research,
Eurasian National University,
Astana, 010008, Kazakhstan
}

\begin{abstract}
Some hypotheses in modern theoretical physics that have not any experimental verification are listed. The goal of the paper is not to criticize or be lawyers any of these hypotheses. The purpose is focus physicists attention on  that now there are too much hypotheses which are not confirmed experimentally.
\end{abstract}


\maketitle

\section{Introduction}

Here we would like present the catalogue of models (theories, hypotheses) in modern physics which are at this time not verified experimentally. It is not possible to construct the full catalogue of such hypotheses but one can  to aspire to make it asymptotically.

The present time is characterized by that there is a considerable quantity of hypotheses which are not checked up experimentally nevertheless are in favor for the last years among physicists. One can say that it is a strange time for physics: since Newton the physics is considered as an experimental science unlike, for example, theology. But now we have many (may be too many) hypotheses which are considered as completed theories and not having any experimental verification. In our world there are not too much absolute statements but one of them is: \textcolor{red}{the physics is the experimental science and any hypothesis should be checked experimentally.} No statements about mathematical (or any another) beauty of the theory can be replaced by the experimental support.

In this notice we do not want to criticize or be lawyers any of these hypotheses. We wish to list only (whenever it is possible) these hypotheses to understand about in what strange (for physics) time we live. This notice is in no case the review of such hypotheses. Therefore our references to corresponding papers will be minimum and certainly are incomplete. In our listing will be resulted both widely known, and little-known hypotheses.

\section{The catalogue}

In this listing the order does not play any role: it could be any other:
\begin{itemize}
\label{list}
	\item string theory;
	\item supersymmetry;
	\item multidimensionality;
	\item loop quantum gravity;
	\item noncommutative geometry;
	\item brane models;
	\item double special relativity;
	\item topological nontriviality;
	\item Hawking radiation;
  \item $F(R)$-gravities; 
	\item nonassociativity.
\end{itemize}

\section{Short description of hypotheses}

Here we would like to give some very short description of every issues in listing \ref{list}.

\subsection{String theory}

The string theory probably is most known hypothesis from the list \ref{list} above (one can see Ref. \cite{stringtext1} as a comprehensive textbook). This theory is known even out of a narrow circle of experts. This hypothesis pretends on the theory of everything. The main idea is very simple and consequently very promising: elementary particles are nothing more than vibrating strings. The difference between particles is the difference between oscillations excited on the string and nothing more. Unfortunately the beauty of this hypotheses does not guarantee the correctness of it.

In Ref. \cite{smolin} one can found full and sufficient analysis of this hypothesis even as a social phenomenon.

\subsection{Supersymmetry}

This hypothesis is beautiful no less than the string hypothesis but it is  less known because the definition of the supersymmetry demands more mathematics. Before supersymmetry bosons and fermions were described by different mathematics. The reason is they have: (1) different statistics, (2)  different quantization technique. For fermions we have so called Pauli principle: not any two fermions may have identical quantum numbers but this principle does not work for bosons. The commutation relationships for fermions and bosons are different: for fermions we have anticommutation relationships but for bosons - commutation relationships. It gives rise to the fact that in Lagrangian we can not mix fermion and boson fields.

Let us note that these fields are different in the sense that: fermions are the matter and bosons transfer the interaction between fermions. For example, electrons (as electric charges) are the sources of  electromagnetic field.

The textbook for the supersymmetry is practically any textbook for quantum field theory, for example, one can see Ref. \cite{ortin}.

\subsection{Multidimensionality}

The hypothesis that our world is multidimensional is very unexpected. After the birth of Einstein's general relativity, the natural question about the dimensionality of the world which we live in appeared. Within the framework of Einstein's gravitational theory, space and time are unified and it allows us to realize that the surrounding world is a 4-dimensional one.

The present stage of the development of multidimensional theories of gravity began with Kaluza's paper \cite{kaluza} where the foundations of modern multidimensional gravitational theories have been laid. The essence of
Kaluza's idea consists in the proposal that the 5-dimensional Kaluza-Klein gravitation is equivalent to 4-dimensional Einstein gravitation coupled to Maxwell's electromagnetism.

One of the most important questions in the multidimensional hypothesis is the question about the unobservability and independence of 4-dimensional quantities on extra dimensions. Now there is two approaches for the resolution of this problem:
\begin{itemize}
	\item it has been considered that in such theories the observable 4-dimensional spacetime appears as a result of compactification of extra dimensions, where the characteristic size of extra dimensions becomes much less than that of 4-dimensional spacetime.
	\item we live on a thin leaf (brane) embedded into some multidimensional space (bulk).
\end{itemize}
Why the multidimensionality is so important ? The answer is: many realistic candidate for a grand unified theory, such as superstring/M-theory,
should be multidimensional by necessity, otherwise, it will contain undesirable physical consequences.

\subsection{Loop quantum gravity}

Loop quantum gravity has matured to a mathematically rigorous candidate quantum field theory of the gravitational field (as a textbook one can see Ref. \cite{thiemann}). The features that distinguish from other quantum gravity theories are: (1) background independence and (2) minimality of structures.

\emph{Background independence} means that this is a non-perturbative approach in which one does not perturb around a given, distinguished, classical metric, rather arbitrary fluctuations are allowed.

\emph{Minimally} means that one explores the logical consequences of bringing together the two fundamental principles of modern physics: general covariance and quantum theory, without adding any experimentally unverified additional structures such as extra dimensions, extra symmetries or extra particles beyond the standard model.

\subsection{Noncommutative geometry}

In a noncommutative geometry the spacetime coordinates $x^\mu$ become noncommuting operators $\hat x^\mu$. The noncommutativity of spacetime is encoded in the commutator
\begin{equation}
	\left[ \hat x^\mu, \hat x^\nu \right] = i \theta^{\mu\nu}
\label{3e-10}
\end{equation}
where $\theta^{\mu\nu}$ is an antisymmetric matrix which determines the fundamental discretization of spacetime (for review one can see \cite{Martin:1996wh}).

\subsection{Brane models}

According to a brane models approach, particles corresponding to electromagnetic, weak and strong interactions are confined on some hypersurface (called a brane) which, in turn, is embedded in some multidimensional space (called a bulk). Only gravitation and some exotic matter (e.g., the dilaton field) could propagate in the bulk. It is supposed that our Universe is such a brane-like object (for review one can see \cite{Maartens:2003tw}).

\subsection{Double special relativity}

According to \cite{dsr}: ``Doubly special relativity \ldots is a modified theory of special relativity in which there is not only an observer - independent maximum velocity (the speed of light), but an observer - independent maximum energy scale (the Planck energy).''

\subsection{Topological nontriviality}

\begin{figure}[h]
  \begin{center}
  \fbox{
    \includegraphics[width=.6\linewidth]{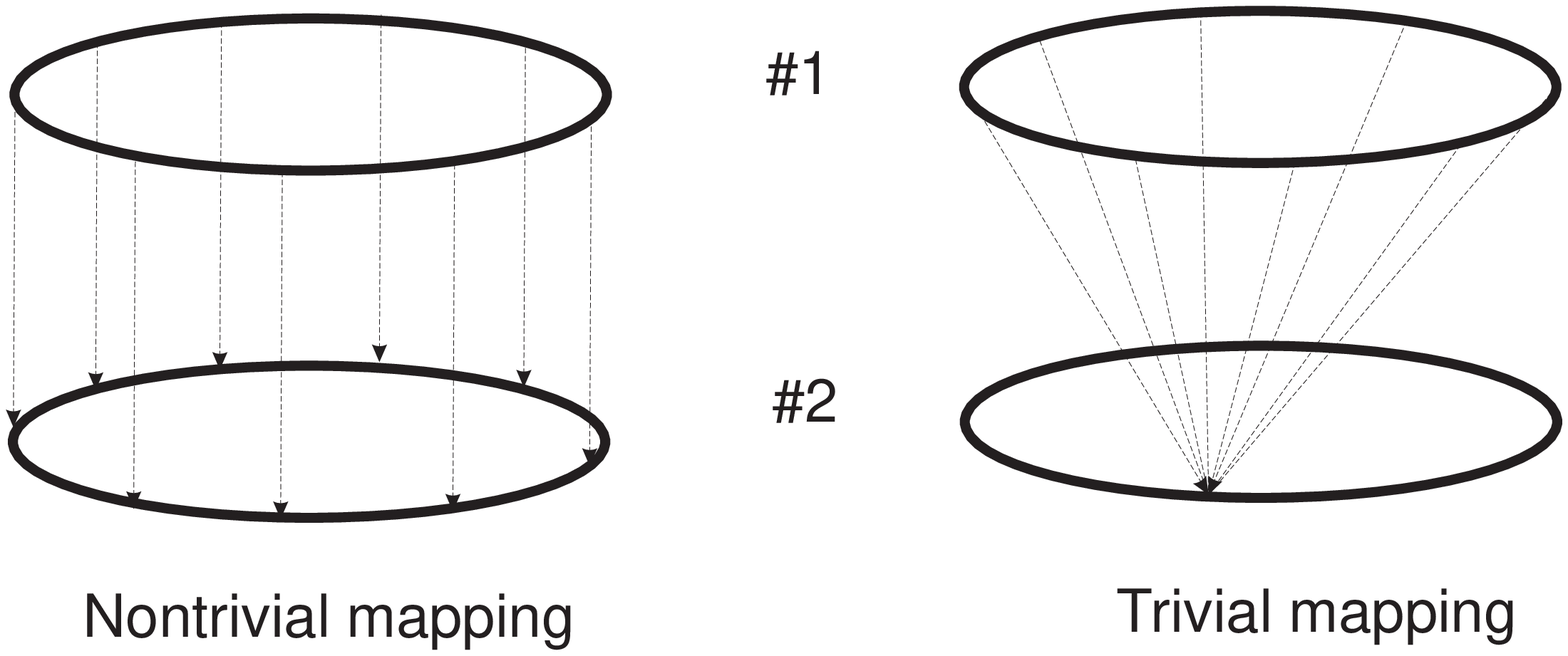}
  }
    \caption{Topological nontrivial and trivial mappings.}
    \label{mapping}
  \end{center}
\end{figure}

The idea about topological nontriviality means that in the Nature these exist objects having topological nontrivial structure. Roughly speaking by topological nontrivial mapping whole object \#1 maps onto whole object \#2. At the same time by topological trivial mapping whole object \#1 maps into one point $\in$ object \#2 (see Fig. \ref{mapping}).

Now we know following topologically nontrivial objects: instanton, monopole and kink. The instanton and monopole are solutions of Yang - Mills equations with very interesting properties (for details one can see practically any quantum field theory textbook, for example: \cite{huang}). The kink is the solution for a nonlinear scalar field.

There are the attempts to apply instantons and monopoles for the resolution confinement problem in quantum chromodynamics. The kink solution is applied for the construction of brane models. There is many other applications for instantons, monopoles and kinks but it is beyond the scope of survey of given paper.

Now we know only one manifestation of topological nontriviality: Abrikosov vortexes in a superconductor. This phenomena is the consequence of the Meissner effect in superconductivity. By this effect a magnetic field is escaped from a superconductor. Abrikosov vortex is a vortex of supercurrent in a type-II superconductor.

\subsection{Hawking radiation}

Hawking radiation is a black body radiation emitted by black holes due to quantum effects. Vacuum fluctuations causes the generation of particle-antiparticle pairs near the event horizon of the black hole. One of the particles falls into the black hole while the other escapes.

\emph{Comments: These calculations are made using perturbative technique. The problem appears if the gravity becomes so strong that perturbative technique can not be applied even for quantum electrodynamics. In this case it becomes  unclear what happened with quantum electromagnetic field in a strong gravitational field. }

\subsection{$F(R)$-gravities}

One of fundamental problems in modern physics is: what is dark energy ? There are many models to understand what is the dark energy: scalar, spinor, 
(non-)abelian vector theory, cosmological constant, fluid with a 
complicated equation of state, higher dimensions and so on. But working in framework of one of the above approaches, it is necessary to introduce the
some extra cosmological components, for example, inflaton, a dark component and dark matter. Unfortunately such a scenarios introduce a new set of problems, for example, coupling with usual matter, (anti-)screening of
dark components during the evolution of the universe, compatibility with standard elementary particle theories and so on. 

Another approach is the gravitational alternative for dark energy: $F(R)$-gravities (for details one can see, for example, \cite{Nojiri:2010wj}). The action for modified gravity is 
\begin{equation}\label{3j-10}
  S_{F(R)}= \int d^4 x \sqrt{-g} \left[ 
  \frac{F(R)}{2\kappa^2} + \mathcal{L}_\mathrm{matter} 
  \right] .
\end{equation}
One can show \cite{Nojiri:2010wj} that some versions of $F(R)$-modified gravities may be consistent with local tests and may provide a qualitatively reasonable unified description of inflation with the dark energy epoch.

\subsection{Nonassociativity}

A nonassociativity is unknown widely hypothesis that in our world there exist a nonassociative structures (for the textbook about the nonassociativity in physics one can see \cite{Okubo}). The people think about: (1) a nonassociative geometry; (2) octonionic version of Dirac equations; (3) octonionic quark structure; (4) a nonassociative decomposition of quantum field operators and nonassociativity in supersymmetric quantum mechanics \cite{Beggs:2005zt}.

\section{Discussion}

In some sense some branches of modern theoretical physics have lost the connection with the experiment. Despite any assurances about beauty of these hypotheses the situation is bad: we cannot name these hypotheses as real physics. The goal of this notice is to remind of this fact and nothing more. The problem consists that nobody can guarantee that these hypotheses are true !

Certainly not all of these hypotheses are very important for the theoretical physics. For example, if the idea about nonassociativity appears to be false it will not lead to any dramatic consequences. But if it will appear that the theory of strings is wrong or in the nature there is no supersymmetry it will lead to the very regrettable consequences.

For physics it will be very bad if in the future the experimental physics cannot give us the answer about the validity of these hypotheses. It may occur, if for example, for this purposes will be necessary unattainable  energy or something like that. It will mean the death of physics as the science: there are questions on which we (because of unattainable energy) cannot answer. Who knows: maybe we live at the natural end of physics. Certainly it is too pessimistic view on physics but we will hope for the best.

From another side we have several fundamental unresolved problems in modern physics. Some of them are: nonperturbative quantization, confinement in quantum chromodynamics, dark energy, dark matter, quantum gravity and so on. Of course this list can vary depending on the point of view of the physicist. For the hypotheses listed in \ref{list} it will be very important to solve even one of these problems. But from another side if one of these problems will be resolved with any another approach it will be very bad for these hypotheses: It means that these hypotheses are not be able to solve these problems. After that the following question appears: for what of problems these hypotheses are necessary ?

\section*{Acknowledgements}

I am grateful to the Research Group Linkage Programme of the Alexander  von
Humboldt Foundation for the support of this research.

\end{document}